\documentclass[runningheads]{llncs}
\usepackage{graphicx}
\usepackage{amsmath}
\begin{document}
\title{Automatic CAD-RADS Scoring using Deep Learning}
\titlerunning{Automatic CAD-RADS Scoring using DL}
%
\author{Felix Denzinger\inst{1,2}
\and
Michael Wels\inst{2} \and 
Katharina Breininger\inst{1} \and
Mehmet A. G\"uls\"un \inst{2} \and
Max Sch\"obinger \inst{2} \and
Florian	Andr\'e \inst{3} \and
Sebastian Bu\ss \inst{3} \and
Johannes G\"orich \inst{3} \and
Michael S\"uhling \inst{2} \and
Andreas Maier \inst{1}
 }
%

\authorrunning{F. Denzinger et al.}
%
\institute{Pattern Recognition Lab, Universit\"at Erlangen-N\"urnberg, Erlangen, Germany \and
 Siemens Healthcare GmbH, Computed Tomography, Forchheim, Germany \and
 Das Radiologische Zentrum - Radiology Center, Sinsheim-Eberbach-Erbach-Walldorf-Heidelberg, Germany
 }
\maketitle              
\begin{abstract}
Coronary CT angiography (CCTA) has established its role as a non-invasive modality for the diagnosis of coronary artery disease (CAD). The CAD-Reporting and Data System (CAD-RADS) has been developed to standardize communication and aid in decision making based on CCTA findings. The CAD-RADS score is determined by manual assessment of all coronary vessels and the grading of lesions within the coronary artery tree.
 
We propose a bottom-up approach for fully-automated prediction of this score using deep-learning operating on a segment-wise representation of the coronary arteries. The method relies solely on a prior fully-automated centerline extraction and segment labeling and predicts the segment-wise stenosis degree and the overall calcification grade as auxiliary tasks in a multi-task learning setup.

We evaluate our approach on a data collection consisting of 2,867 patients.
On the task of identifying patients with a CAD-RADS score indicating the need for further invasive investigation our approach reaches an area under curve (AUC) of 0.923 and an AUC of 0.914 for determining whether the patient suffers from CAD. This level of performance enables our approach to be used in a fully-automated screening setup or to assist diagnostic CCTA reading, especially due to its neural architecture design -- which allows comprehensive predictions.

\end{abstract}
\section{Introduction}
Coronary Artery Disease (CAD), which may lead to major adverse events like cardiac infarction or significantly decrease quality of life in the form of coronary ischemia, remains the most common cause of death \cite{mendis15}.
Most kinds of CAD result from atherosclerotic plaque deposits aggregating in the vessel wall creating a stenosis, hence narrowing the vessel and obstructing the blood flow. The plaque lesions are categorized by the degree of stenosis into no (0\%), minimal (1-24\%), mild (25-49\%), moderate (50-69\%), severe stenosis (70-99\%), and occluded vessel (100\%) \cite{cury16}.\\
Coronary CT Angiography (CCTA) is a common non-invasive rule-out modality for CAD due to its high negative predictive value. In order to standardize communication and guide patient management, the CAD-RADS score based on above mentioned stenosis grades was introduced \cite{cury16}. It ranges between 0 and 5 and is strongly influenced by the degree of the severest stenosis within a patient. 
Additionally, this score is influenced by the location of the lesion and includes qualitative assessments based on the experience of the physician, especially in edge-cases.  \\ 
From a high-level perspective for the case of stable CAD, the resulting patient management decision can be divided into three options: the patient has no CAD and does not need any treatment in the direction of CAD (0), the patient has a non-obstructive CAD (1-2) without need for further investigation, or the patient has an obstructive CAD and should undergo a further functional investigation or direct intervention (3-5). \\
Therefore, at least these clinical questions need to be answered by an assisting image analysis tool: in the rule-out case, the CAD-RADS 0 score needs to be differentiated from 1-5, and in the hold-out case, the CAD-RADS scores 0-2 need to be differentiated from 3-5. However, prediction on an even finer scale is necessary when the exact required action needs to be identified. \\
In clinical practice, the assessment of the CAD-RADS score is cumbersome, since the whole coronary tree needs to be assessed and the severest lesion is graded manually based on experience and eyeballing, which is prone to error. Therefore, approaches to ease the workflow and help to detect and grade stenotic lesions have been developed in recent years. 
Previous approaches focus on detection and quantification of stenoses and are based on the segmentation of the entire coronary tree \cite{kiricsli13,sankaran2016hale}, which is time consuming and often needs manual correction \cite{wels16}.\\
Recently, deep-learning approaches \cite{maier2019gentle} without the need for a prior segmentation were introduced \cite{candemir19,tejero19,zreik18b}. These methods operate on multi-planar reformatted (MPR) image stacks which are extracted by interpolating orthogonal planes for each centerline point of the vessel. Approaches for this task include a recurrent convolutional neural network (RCNN) \cite{zreik18b}, a 2D texture-based multi-view \cite{tejero19}, and a 3D CNN approach \cite{candemir19}. A 2D CNN approach, which classifies the whole CCTA volume scaled down and placed in a 2D grid, is described in \cite{muscogiuri}, but might have optimistic results since the training and test splits are described not to be patient-wise.\\
However, most of the above approaches have the disadvantage of determining the patient score based on single lesions, again introducing a large amount of potential error sources, with no global context incorporated into the decision. \\ 
To overcome these pitfalls, we propose a bottom-up approach to directly predict the patient-level CAD-RADS score using a deep-learning based approach that leverages a task-specific hierarchical data representation building up on the coronary tree segments as defined by the American Health Association (AHA) norm.  
By having the segment-wise stenosis degree as an additional output and by utilizing a global max pooling operation, which identifies the most relevant features across the whole coronary tree, the network is designed to be comprehensive. Additionally, since all steps in the workflow of our approach can be automated, it can be used for patient screening as well as a preprocessing utility to ease and speed up the clinical workflow.

\section{Data}\label{sec:label}
We train and evaluate our methods on a data set consisting of CCTA scans from 2,867 patients collected at a single site. \\
For each patient, labels regarding the stenosis degree were given on a segment-level as no-stenosis, minimal, mild, moderate, severe or occluded with frequencies of 3,625, 34,889, 4,565, 2,324, 722 and 70 
and on patient-level with frequencies of 53, 940, 861, 611, 352 and 50. Furthermore, the CAD-RADS score was annotated on the patient-level with categories 0-5 with frequencies of 436, 584, 873, 568, 348 and 58 \cite{cury16}. The difference between the patient-wise stenosis degree and the CAD-RADS score can be explained by edge-cases and is especially severe in the CAD-RADS 0 case, since lesions with very minor wall irregularities were classified as minimal according to literature~\cite{cury16}. Additionally, for a subset of 2,828 patients, the Agatston scores were annotated based on additional calcium scoring scans, which were utilized in a binned version according to Rumberger et al.~\cite{rumberger} as no, minimal, mild, moderate and severe calcifications with frequencies of 911, 317, 649, 491 and 460.\\
The data collection did not include patients with stents or bypass grafts. 
It was split into two parts with two thirds (1,899) used for training and one third (968) used for testing. 

\section{Methods}
\subsubsection{Preprocessing}
For each patient, centerlines are automatically extracted using the algorithm described by Zheng et al.~\cite{zheng13} and assigned to the AHA segments~\cite{gulsun14}. The extracted AHA segment centerlines are used to create MPR image stacks, which are then resized to the mode segment length resulting in a subvolume of size $128\times32\times32$ for each segment according to Denzinger et al. \cite{denzinger20}. Subsequently, the Hounsfield Unit (HU) value range is clipped between -324 and 1,176 HU and normalized to a value range of $[0,1]$. In order to focus on the more important sections and prevent error propagation from mislabeled AHA-segments, we only select a subset of AHA-segments (RCA\_p, RCA\_m, RCA\_d, LM, LAD\_p, LAD\_m, LAD\_d, LAD\_D1, CX\_m, CX\_d, RAMUS), which were more robustly labeled according to G\"uls\"un et al.~\cite{gulsun14}. We confirmed to reach similar performance with this subset compared to utilizing all segments in preliminary experiments.\\
\subsubsection{Neural Architecture Design}
\begin{figure}[t]
    \centering
    \includegraphics[width=0.95\textwidth]{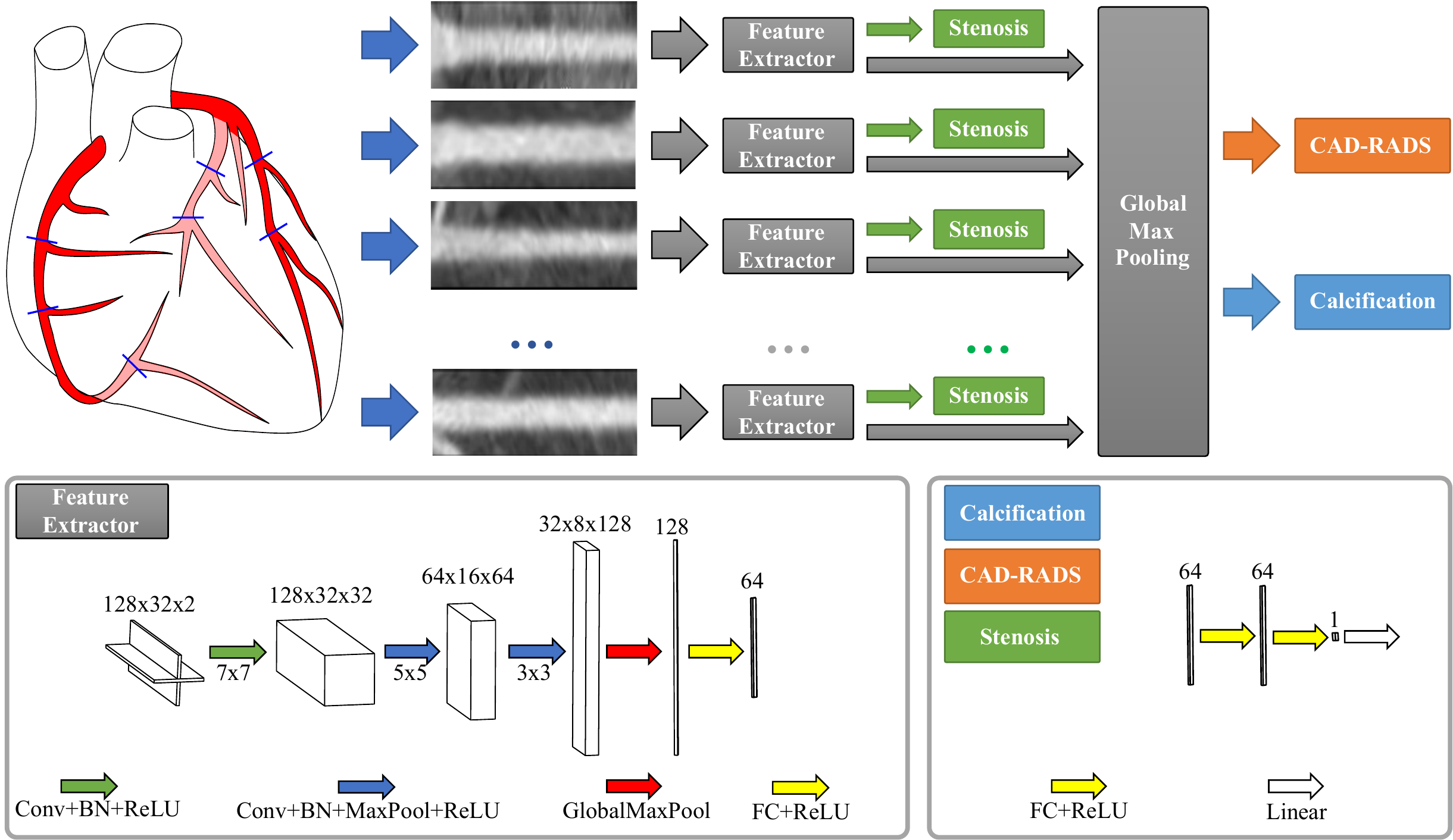}
    \caption{Model overview 
    (Conv = 2D Convolutional Layer; BN = Batch Normalization Layer; MaxPool = 2D MaxPooling Layer; ReLU = Rectified Linear Unit; FC = Fully-Connected Layer).}
    \label{fig:model}
\end{figure}

The general workflow of the proposed method is outlined in Figure~\ref{fig:model}. Since our classes are ordered, we reformulate our classification task as a regression problem. This carries the benefit that misclassifications are penalized stronger depending on the class distance, which is convenient since misclassifications between neighboring classes are not as severe for our task.  
In order to reformat the whole coronary tree in a reasonable representation for neural network training, we divided the whole coronary tree into its sub-segments and extracted straightened MPR volumes. Since we assume all segments to be able to contribute equally, we utilize a feature extractor block with shared weights across all segments to extract spatial features. The feature extractor blocks work on a 2.5D representation utilizing a simple convolutional neural network (CNN)~\cite{denzinger20}. This architecture choice is motivated by the fact that we strived for simple building blocks to reduce the overall computational effort. Furthermore, we validated in prior experiments that adding additional views or having a feature extractor block similar to the method of \cite{zreik18b} did not improve the performance. Since we do not want our model to depend on the location of the stenosis within one segment we choose to decouple the spatial features using a global max pooling operation. A fully-connected layer is used as the last layer of the feature extractor block in order to weight and combine the features such that our different targets can influence each other in the multi-task learning setup. The output of the feature extractor block is then either processed by a stenosis regression block with shared weights across segments to predict the stenosis degree of each segment or the maximum feature responses across all segments are extracted by a global max pooling layer. These global maximum feature responses are then fed into two further regression blocks for the CAD-RADS and calcification score prediction. This architecture choice is motivated by the definition of the CAD-RADS score as being heavily influenced by the severest lesion. Furthermore, the use of global max pooling allows the network to be more comprehensible since the regions with the highest activations as determined by the network can be displayed to the physician. \\
\subsubsection{Evaluation}
In order to evaluate the effectiveness of the use of multi-task learning, we evaluated our approach on three different configurations: directly regressing the CAD-RADS score (CAD-RADS), additionally regressing the segment-wise stenosis scores (CAD-RADS + $\asymp$) and also regressing the calcification score (CAD-RADS + $\asymp$ + Ca). Furthermore, to verify whether the global context introduced by our architecture improves the performance, we also evaluate the combination of the feature extractor block and the stenosis regression block with the severest prediction being propagated to the patient-level (Patient-level $\asymp$), which is as close as we can get to related work algorithms with our given labels. \\
The training set is split into five folds of actual training and validation data (80$\,\%$/20$\,\%$). The model with the overall lowest loss on the validation set is used as a checkpoint for later evaluation.
We choose the Adam optimizer with a learning rate of 0.0001, a batch size of 32 and mean squared error loss for all targets. Furthermore, we utilize data augmentation in the form of rotations around the centerline and minor shifts in x and y direction. In all experiments involving the segment-wise stenosis grade, the feature extractor block is pretrained on the stenosis grade on segment-level before getting integrated into the full model. This is done to condition the feature extractor block towards learning relevant features for the prediction of the stenosis degree. 
In order to convert our regressed predictions back into classes, we enforce the binned predictions to have the same class distribution as the ground truth labels. The thresholds used for this are calculated on the training set and propagated to the test set. 

\section{Results} 
As mentioned in Section~1, most reference approaches perform the classification of the severeness on a per-lesion-level with only Zreik et al. \cite{zreik18b} performing an evaluation on the patient-level. However, the severest lesion per patient is not equivalent to the CAD-RADS score and differs especially often in the CAD-RADS 0 case (see Section~2), hence complicating a direct comparison. 
\subsubsection{CAD-RADS Performance}
\begin{figure}
    \centering
    \includegraphics[width=.6\textwidth]{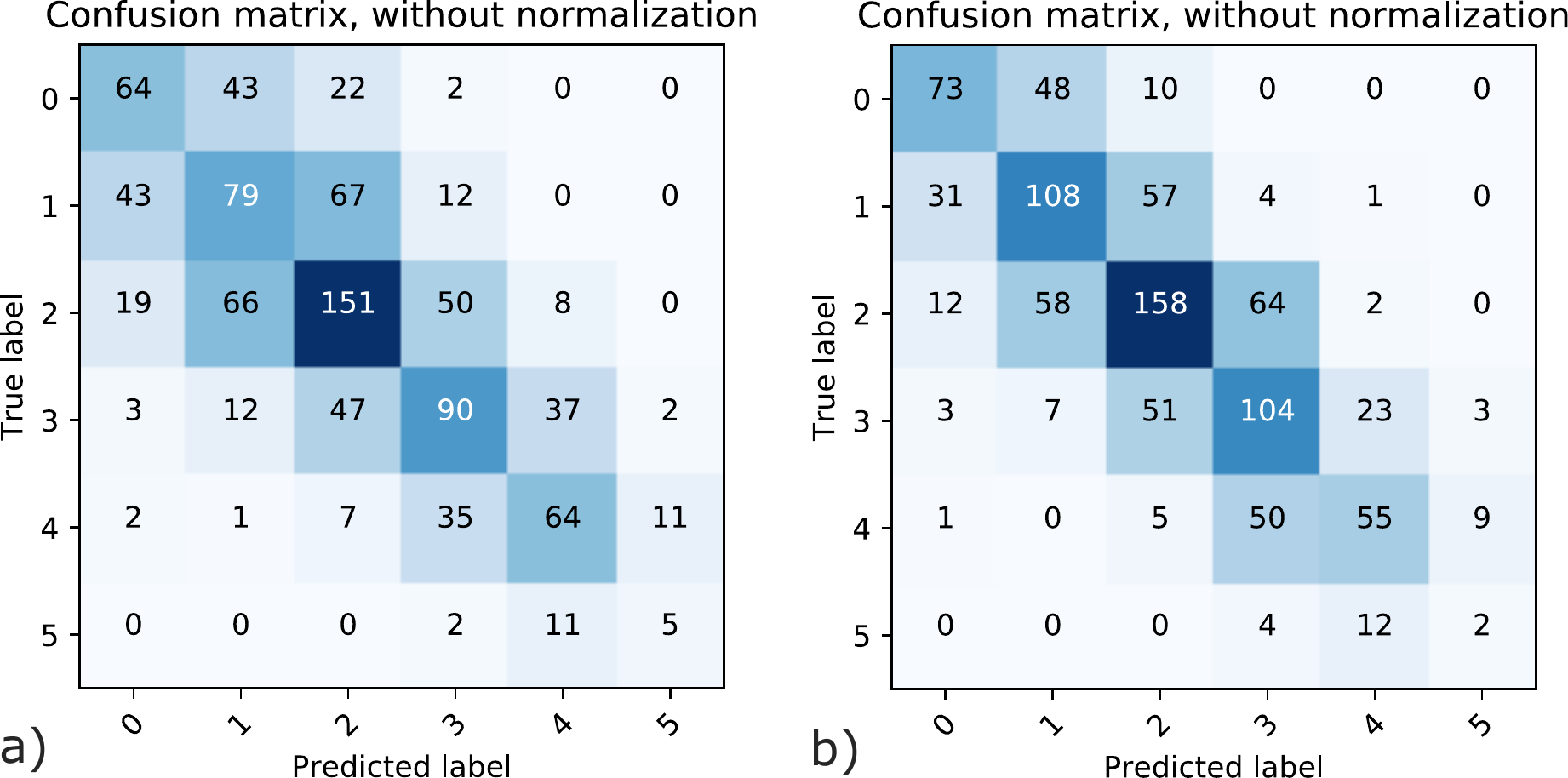}
    \caption{Example confusion matrices of a single fold on the target of predicting the CAD-RADS using the maximum segment-wise prediction (a) and our proposed multi-task learning procedure (b).}
    \label{fig:confusion}
\end{figure}

Before analyzing the clinical tasks at hand (rule-out/hold-out), we want to analyze the performance of our approach under different configurations for all six classes. 
Results for our baseline (severest lesion score as patient score) approach and our full model are given in Fig.~\ref{fig:confusion} and Tab.~\ref{table:overall}. 
By leveraging multi-task learning we are able to boost the performance of our approach incrementally (Tab.~\ref{table:overall}) from an accuracy of 0.810 to 0.840. While the baseline approach performs better compared to direct CAD-RADS scoring without auxiliary targets, we manage to outperform it in the multi-task setup. The biggest performance difference in comparison to the baseline are the lower CAD-RADS scores since in these cases overestimation of single-segment stenoses degrees are especially severe. As displayed in Fig.~\ref{fig:confusion}, the hardest class to identify was CAD-RADS 5. An explanation for this is the fact that the centerline extraction fails in the case of occluded vessels. Our method has a low specificity due to the high class imbalance for the single class metrics. 
Apart from this, most misclassifications are within one class distance, especially in our multi-task learning setup, which is a good feature with respect to the confidence in the network decision.
\begin{figure}
    \centering
    \includegraphics[width=.98\textwidth]{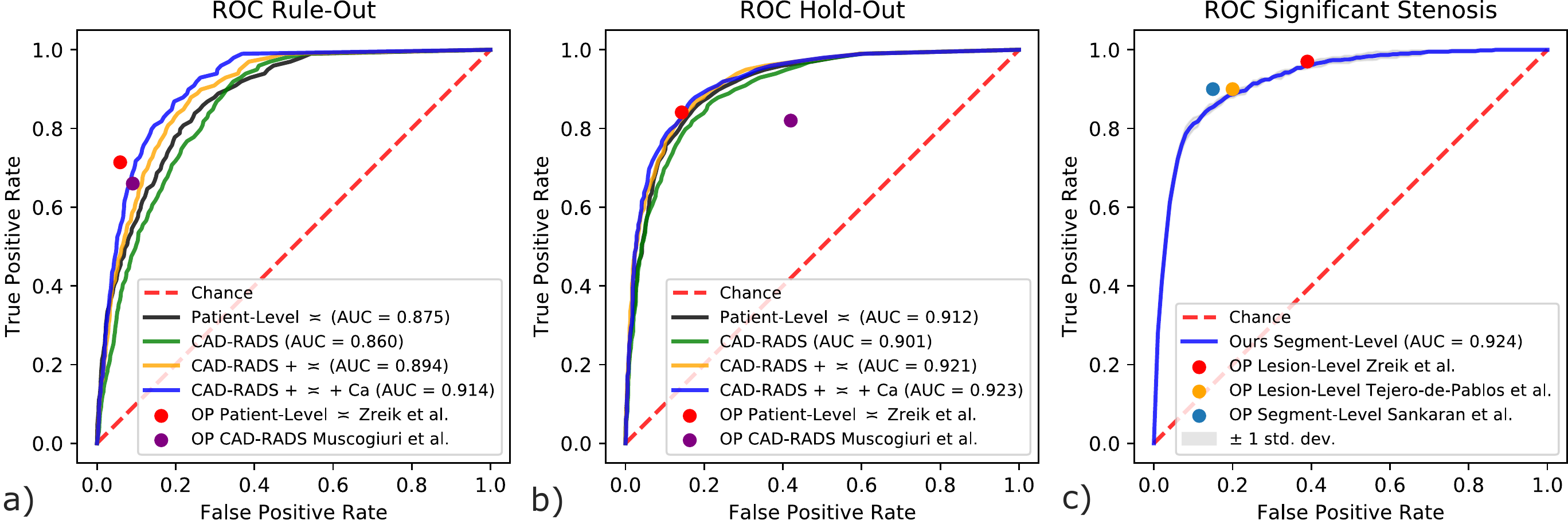}
    \caption{Results: a) Mean receiver operating characteristic (ROC) curves for the rule-out case. The operating points (OP) of Zreik et al. \cite{zreik18b} and Muscogiuri et al. \cite{muscogiuri} refer to metrics calculated on their data set with Zreik et al. operating on the related task of classifying the severest stenosis degree.  b) Mean ROC curves for the hold-out case. c) Mean ROC curve for the classification of significant stenoses. Prediction in Zreik et al.~\cite{zreik18b} and Tejero-de-Pablos et al.~\cite{tejero19} is performed on a per-lesion level and Sankaran et al.~\cite{sankaran2016hale} utilize the vessel segmentation as additional preprocessing.}
    \label{fig:AUCsAuxtask}
\end{figure}
	\begin{table}
		\begin{minipage}{\textwidth}
			\centering
			\caption{Mean performance on the six class problem of the baseline approach and the three different multi-task learning network configurations ($\asymp$ = segment-wise stenosis grade; Ca = patient-wise calcification grade; MCC = Matthews Correlation Coefficient).}\label{table:overall}
			\begin{tabular}{ l | l | l | l | l }
				\hline
				Approach/Metric  & Accuracy & Sensitivity & Specificity & MCC\\
				\hline
				\hline
				Patient-level $\asymp$  & 0.825 &	0.895 &  0.476 &	0.371 \\ 
				\hline
				CAD-RADS & 0.810 &	0.886 &  0.430 &	0.316 \\ 
				\hline
				CAD-RADS + $\asymp$ & 0.832 &	0.899 &  0.496 &	0.395\\ 
				\hline 
				CAD-RADS + $\asymp$ + Ca & \textbf{0.840} &	\textbf{0.904} &  \textbf{0.520} &	\textbf{0.424}\\	
				\hline
			\end{tabular}
		\end{minipage}
	\end{table}
\subsubsection{Rule-out}
On the task of classifying whether a patient suffers from CAD, we see incremental improvements in the performance of our method with each auxiliary target from an AUC of 0.860 to 0.894 to 0.914 (Fig.~\ref{fig:AUCsAuxtask}a and Tab.~\ref{table:ruleout}). The performance boost of utilizing the calcification grade can be explained by the fact that patients without CAD should not exhibit any calcifications in the coronary arteries. 
Also the baseline approach of propagating the severest segment-prediction to the patient-level only reaches an AUC of 0.875 compared to the 0.914 of our full model.
Furthermore, there is a severe gap between sensitivity and specificity due to class imbalance. However, as the ROC curve (Fig.~\ref{fig:AUCsAuxtask}a) indicates an operating point with both sensitivity and specificity above 0.800 -- which is often times required in a clinical setting -- can be selected. 
\begin{table}
		\begin{minipage}{\textwidth}
			\centering
			\caption{Results for the rule-out case (predicting CAD-RADS 0 vs 1-5). Results of Zreik et al. \cite{zreik18b} refer to the related but different task of predicting the severest stenosis degree on a different data set (abbreviations as in Table 1).}\label{table:ruleout}
			\begin{tabular}{ l | l | l | l | l | l | l }
				\hline
				Approach/Metric & Patients & AUC & Accuracy & Sensitivity & Specificity & MCC\\
				\hline
				\hline
				Patient-level $\asymp$  & 955 & 0.875 &	0.865 &  0.508 &	0.921&	0.430 \\ 
				\hline
				CAD-RADS  & 955 & 0.860 &	0.849 &  0.489 &	0.907&	0.384 \\ 
				\hline
				CAD-RADS + $\asymp$  & 955 & 0.894 &	0.875 &  0.510 &	0.933&	0.456 \\ 
				\hline
				CAD-RADS + $\asymp$ + Ca &  955 & \textbf{0.914} &	\textbf{0.888} &  \textbf{0.532} &	\textbf{0.945}&	\textbf{0.504}\\	
				\hline
				\hline
				Zreik et al. \cite{zreik18b}& 65 & -  &	0.892 & 0.714  &	0.941&	0.674\\	
				\hline
				Muscogiuri et al. \cite{muscogiuri}& 284 & 0.89&	0.863  &	0.660 & 0.909  &	0.558\\	
				\hline
			\end{tabular}
		\end{minipage}
		
	\end{table}
	
\subsubsection{Hold-out}
In the hold-out case, the use of auxiliary tasks did not boost the performance as much as for the other targets (Fig.~\ref{fig:AUCsAuxtask}b and Tab.~\ref{table:holdout}), with the biggest gain caused by adding the segment-wise stenosis degree. However, we outperform our baseline with an AUC, accuracy and MCC of 0.923, 0.860 and 0.692. 
\begin{table}
		\begin{minipage}{\textwidth}
			\centering
			\caption{Results for the hold-out case (predicting CAD-RADS 0-2 vs 3-5). (abbreviations as in Table~\ref{table:overall})}\label{table:holdout}
			\begin{tabular}{ l | l | l | l | l | l | l }
				\hline
				Approach/Metric & Patients & AUC & Accuracy & Sensitivity & Specificity & MCC\\
				\hline
				\hline
				Patient-level $\asymp$  & 955 & 0.912 &	0.850 &  0.885 &	0.781&	0.666 \\ 
				\hline
				CAD-RADS & 955 & 0.901 &	0.838 &  0.879 &	0.759&	0.640 \\ 
				\hline
				CAD-RADS + $\asymp$  & 955 & 0.921&	0.858&  \textbf{0.895} &	0.787&	0.684 \\ 
				\hline
				CAD-RADS + $\asymp$ + Ca& 955 & \textbf{0.923} &	\textbf{0.860}&  0.891 &	\textbf{0.802}&	\textbf{0.692}\\	
				\hline
				\hline
				Zreik et al. \cite{zreik18b}&65 & -  &	0.846 &  0.841 &	0.857&	0.671\\	
				\hline
				Muscogiuri et al. \cite{muscogiuri}& 284 & 0.78&	0.711  &	0.822 & 0.583  &	0.420\\	
				\hline
			\end{tabular}
		\end{minipage}
		
	\end{table}
\subsubsection{Auxiliary Targets}
For the target of predicting the stenosis degree on a segment-wise level, we reach results comparable to state-of-the-art methods when looking at the binary case of predicting significant stenosis ($\,>$50\,\%) (Fig.~\ref{fig:AUCsAuxtask}c). It should be noted that competing methods are evaluated on different data sets and use labels on lesion-level with defined start and end points, which require a remarkable amount of effort for annotation. Furthermore, our performance on this level enables that segments with the highest score are highlighted in order to aid physicians in their decision making process. \\
On the task of predicting our calcification grade (as defined in Section~2) we are able to reach a mean accuracy of 0.878. 

\section{Conclusion}
In clinical practice, a standardized way to report CAD from CCTA scans is the CAD-RADS score.
To the best of our knowledge -- this work presents and evaluates the first approach to directly predict the six class CAD-RADS score using a deep-learning based algorithm. 
By leveraging two auxiliary tasks -- the prediction of the segment-wise stenosis grade and a patient-wise calcification grade -- we boosted the performance of our method. The method only relies on a prior centerline extraction and AHA segment label but not on the segmentation of the coronary tree, which is time-consuming to obtain and may need manual correction. Our approach is able to robustly identify patients suffering from CAD (AUC 0.914) or requiring further clinical investigation (AUC 0.923). Segments with severe lesions can be identified by our approach due to the neural architecture design and since we predict segment-wise stenosis with the same network. We validated our approach on a data set of 2,867 patients, a data set considerably larger compared to what has been reported in related work. \\
Still, the used 2.5D data representation of the single segments may omit some 3D information. We expect this to be successfully addressed by using test augmentation or utilizing additional views in future work. Within this study, it was not possible to apply algorithms defined in related work to our data set, since our stenosis degree labels were segment-wise and not on a per lesion-level. Still, with our experimental design we address this issue in order to allow for a fair comparison. Furthermore, the definition of CAD-RADS also includes report modifiers related to high-risk plaques, stents and bypass grafts  which will be addressed in future work.
%
%
\subsubsection*{Disclaimer}
	The methods and information here are based on research and are not commercially available.

\bibliographystyle{splncs04}
\bibliography{samplepaper}
\end{document}